# Nontrivial Berry Phase in Magnetic BaMnSb$_2$ Semimetal

Silu Huang[a], Jisun Kim[a], W. A. Shelton[b], E. W. Plummer[a], Rongying Jin[a,1]

[a]Department of Physics and Astronomy, Louisiana State University, Baton Rouge, LA 70803, USA

[b]Department of Chemical Engineering, Louisiana State University, Baton Rouge, LA 70803. USA

[1]To whom correspondence may be addressed.

**Corresponding Author**

**Rongying Jin**

Address: 229-B Nicholson Hall, Tower Dr., Department of Physics & Astronomy, Louisiana State University, Baton Rouge, LA 70803-4001, USA

Phone Number: +1-225-578-0028

E-mail: rjin@lsu.edu

**ABSTRACT**

The subject of topological materials has attracted immense attention in condensed-matter physics, because they host new quantum states of matter containing Dirac, Majorana, or Weyl fermions. Although Majorana fermions can only exist on the surface of topological superconductors, Dirac and Weyl fermions can be realized in both 2D and 3D materials. The latter are semimetals with Dirac/Weyl cones either not-tilted (type-I) or tilted (type-II). Although both Dirac and Weyl fermions have massless nature with the nontrivial Berry phase, the formation of Weyl fermions in 3D semimetals require either time-reversal or inversion symmetry breaking to lift degeneracy at Dirac points. Here, we demonstrate experimentally that canted antiferromagnetic $BaMnSb_2$ is a 3D Weyl semimetal with a 2D electronic structure. The Shubnikov-de Hass oscillations of the magnetoresistance give nearly zero effective mass with high mobility and the nontrivial Berry phase. The ordered magnetic arrangement (ferromagnetic ordering in the *ab* plane and antiferromagnetic ordering along the *c* axis below 286 K) breaks the time-reversal symmetry, thus offering us an ideal platform to study magnetic Weyl fermions in a centrosymmetric material.

**Significance Statement**

Among topological materials, experimental study of topological semimetals that host Dirac/Weyl fermions has just begun, even though these topological concepts were proposed nearly a century ago by Dirac and Weyl. Our work shows magnetic semimetal BaMnSb$_2$ exhibits nearly zero-mass fermions with high mobility and a nontrivial Berry phase. What is unique is the magnetic ordering, indicating the system is Weyl type due to time-reversal symmetry breaking. Theory shows that the spin order is very fragile, so it is expected that the application of magnetic field or uniaxial pressure could drive the material to be a type-II Weyl semimetal. In this material Mn mainly controls the magnetic structure, whereas Sb in only one lattice site is responsible for the Dirac behavior.

The observation of the quantum Hall effect has led to the discovery of new phases associated with topological ordering. In the past decade, topological materials have emerged as a new frontier of condensed matter physics, due to new physical concepts and potential applications. Theory has played a major role in predicting new topological materials, which have been realized experimentally. Topological insulators, which are metallic at the surface but insulating in bulk (for a review, see ref. 1), have been extensively investigated. Surface states in these materials can be described by the Dirac equation with the Fermi surface formed by Dirac points. Dirac fermions are effectively massless because their dispersion is linear in energy. If the bulk is superconducting, the surface hosts Majorana fermions (for a review, see ref. 2). More recently, a new class of topological materials, namely, Dirac or Weyl semimetals, has appeared. Such topological semimetals are characterized by the presence of electron and hole pockets touching with either degeneracy (Dirac) or nondegeneracy (Weyl). If both time-reversal and inversion symmetries are preserved, the system is a Dirac semimetal (3). If either time-reversal or inversion symmetry is broken, the Dirac points split and turn into Weyl points, making the system a Weyl semimetal (3). If both time-reversal and inversion symmetries are broken, the system may become a Weyl superconductor (4).

The above framework makes the search of topological semimetals possible. For example, Dirac semimetals should have centrosymmetry to preserve the inversion symmetry. However, a noncentrosymmetric crystal structure would favor Weyl semimetal configuration (5). If a system is centrosymmetric and magnetically ordered, whether time-reversal symmetry is broken would determine the nature of semimetals. In recent experimental reports, magnetic $XMnPn_2$ (X = alkali-earth or rare-earth elements,

Pn = Sb, Bi) system is characterized as either Dirac (6-11) or Weyl semimetals (12,13). Because these materials are centrosymmetric and magnetically ordered at elevated temperatures (6-12), the key is to find if the time-reversal symmetry is broken by specific magnetic arrangement. Among this class of materials, BaMnSb$_2$ is very interesting because its *spin order may not be centrosymmetric*, and it has the lowest reported effective mass (~ 0.052$m_0$) (11). In this article, we report quantum transport of in-plane, out-of-plane, and Hall resistivities for high-quality BaMnSb$_2$ single crystals over a wide temperature and field range. Although first-principle calculations show the existence of Dirac points and low density of states near the Fermi energy, it indicates from our experiment that BaMnSb$_2$ satisfies all of the criteria for being a Weyl semimetal due to its magnetic arrangement: broken time-reversal symmetry and nontrivial Berry phase.

**Results and Discussion**

Figure 1*A* depicts the crystal structure of BaMnSb$_2$, consisting of alternately stacked MnSb, Ba, Sb, and Ba layers. Due to its tetragonal structure, the Sb layer forms the square lattice (14). According to theoretical calculations (12,15), such square lattice is critical to the formation of Dirac/Weyl points. We have successfully grown single crystalline BaMnSb$_2$ via the floating-zone technique as shown in Fig. 1*B* (see also *Materials and Methods*). Electron microprobe analysis shows that the composition of single crystals used in this study can be expressed as BaMn$_{1-\delta}$Sb$_2$ with $0 < \delta < 0.05$. According to our x-ray diffraction analysis on the single crystals, BaMn$_{1-\delta}$Sb$_2$ crystallizes in a tetragonal structure (I4/mmm), with $a = b = 4.556$ Å and $c = 24.299$ Å. These values are consistent with that obtained previously (11,14,15).

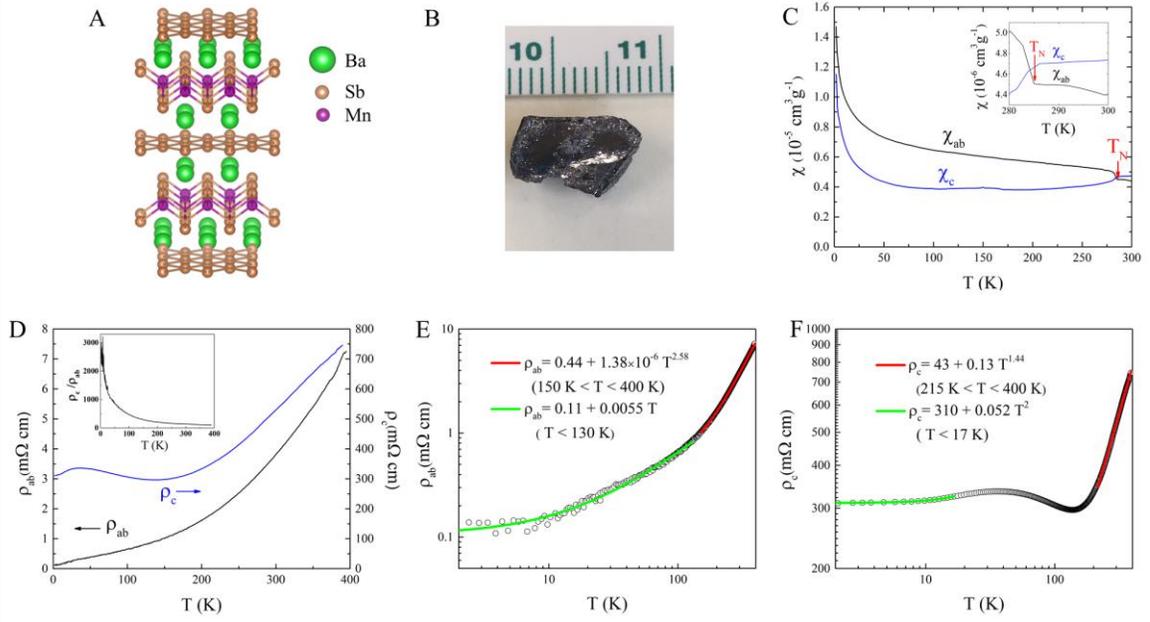

**Figure 1**: (A) Schematic crystal structure of BaMnSb$_2$. (B) Image of a single crystal. (C) Temperature dependence of magnetic susceptibilities along the *ab* plane ($\chi_{ab}$) and *c* direction ($\chi_c$). *Inset* shows magnetic susceptibilities near the transition temperature $T_N$. (D) Temperature dependence of electrical resistivities along the *ab* plane ($\rho_{ab}$) and *c* direction ($\rho_c$). *Inset* shows the temperature dependence of $\rho_c/\rho_{ab}$. The data are replotted in logarithmic scales for (E) $\rho_{ab}(T)$ and (F) $\rho_c(T)$.

Given the centrosymmetric crystal structure, it is essential to characterize the magnetic properties to ascertain if time-reversal symmetry is broken in BaMnSb$_2$. Fig. 1*C* shows the temperature dependence of the magnetic susceptibilities $\chi_{ab}$ and $\chi_c$ measured by applying magnetic field at 0.1 T. With decreasing temperature, $\chi_{ab}$ slowly increases, whereas $\chi_c$ is more or less constant for $T > 286$ K. Below 286 K, $\chi_{ab}$ suddenly increases, whereas $\chi_c$ decreases (see Fig. 1*C*, *Inset*). This indicates that there is a magnetic transition at $T_N = 286$ K, with ferromagnetic (FM) character in the *ab* plane and antiferromagnetic (AFM) behavior along the *c* direction. Similar magnetic response has been reported for YbMnBi$_2$ (12), Sr$_{1-y}$Mn$_{1-z}$Sb$_2$ (13), and BaMnSb$_2$ (11), but our result shows $\chi_{ab} > \chi_c$ below $T_N$, whereas ref. 11 reports that $\chi_c > \chi_{ab}$.

The temperature dependence of the in-plane ($\rho_{ab}$) and c-axis ($\rho_c$) resistivites between 1.8 and 400 K is shown in Fig. 1D. Although $\rho_{ab}$ decreases smoothly with decreasing temperature, $\rho_c$ has much larger magnitude with a bump located near 50 K. The resistivity anisotropy $\rho_c/\rho_{ab}$ increases dramatically with decreasing temperature from ~90 at 400 K to ~1250 at 2 K (Fig. 1D, *Inset*), much larger than the anisotropy seen in magnetic susceptibility (Fig. 1C). There is no sign for the magnetic transition at $T_N$ in either $\rho_{ab}$ or $\rho_c$. Qualitatively, the profile of our $\rho_{ab}(T)$ and $\rho_c(T)$ is similar to that seen in Sr$_{1-y}$Mn$_{1-z}$Sb$_2$ (13), whereas the bump is located at a much higher temperature for the latter. Again, both $\rho_{ab}(T)$ and $\rho_c(T)$ are rather different from that reported in ref. 11, even though the resistivity anisotropy is comparable.

Quantitatively, we find that $\rho_{ab}$ shows power-law temperature dependence, i.e., $\rho_{ab} = A_{ab} + B_{ab}T^\alpha$ ($A_{ab}$, $B_{ab}$, and $\alpha$ are constants). As demonstrated in Fig. 1E, we can fit our data in two different temperature ranges, above or below 150 K, resulting in two sets of $A_{ab}$, $B_{ab}$, and $\alpha_{ab}$ values (shown in Fig. 1E). Above ~ 150 K, the $\rho_{ab}$ data can be fit using $A_{ab}$ = 0.44 mΩ cm, $B_{ab}$ = 1.38×10$^{-6}$ mΩ cm K$^{-2.58}$, and $\alpha$ = 2.58 (see the red curve in Fig. 1E). Below ~ 150 K, $\rho_{ab}(T)$ shows much weaker temperature dependence with linear behavior ($\rho_{ab} \propto T$) (see the green curve in Fig. 1E).

Although there is a small bump near 50 K, $\rho_c$ away from this temperature displays power-law temperature dependence, i.e., $\rho_c = A_c + B_cT^\beta$ ($A_c$, $B_c$, and $\beta$ are constants). Shown in Fig. 1F is the logarithmic plot of $\rho_c(T)$ with the fitting. Above ~ 215 K, $\rho_c(T)$ can be fit with $A_c$ = 43 mΩ cm, $B_c$ = 0.13 mΩ cm K$^{-1.44}$, and $\beta$ = 1.44 (see the red curve in Fig. 1*F*). Below ~ 15 K, $\rho_c(T)$ shows more or less quadratic temperature dependence ($\rho_c \propto T^2$) (see the green curve in Fig. 1*F*).

The quantitative analysis of $\rho_{ab}(T)$ and $\rho_c(T)$ indicates that BaMnSb$_2$ is not a conventional metal: the observed power-law temperature dependence is inconsistent with either electron-phonon and/or electron-electron scattering mechanisms. According to ref. 13, the temperature profile of $\rho_{ab}$ and $\rho_c$ from Sr$_{1-y}$Mn$_{1-z}$Sb$_2$ depends strongly on sample magnetism. It is likely that the unusual temperature dependence of resistivity is the consequence of charge-spin scattering, even though it is not reflected at $T_N$.

To understand the unusual electronic and magnetic properties of BaMnSb$_2$, we have performed relativistic first-principles calculations including spin-orbit coupling with an on-site Hubbard-like Coulomb interaction term $U$ = 5 eV. Fig. *2A* shows the Brillouin zone (BZ) for BaMnSb$_2$, where the z direction is perpendicular to the *ab* plane. Fig. 2*B* is the calculated band structure for relevant directions in the BZ. Fig. 2*B*, *Bottom*, is for $k_z$ = 0 in the diagonal direction $\Gamma \to M \to X$, showing a Dirac cone at ~40% of the way to the zone boundary. Fig. 2*B*, *Top*, is for $Z \to A \to R$ in the same direction in $k_\parallel$, but with $k_z$ = Z. Fig. 2*B*, *Middle*, is the same direction but with $k_z$ = Z/2. Note that the electronic structure is 2D, especially the Dirac cone, which is derived from Sb in the square lattice (Fig. 2*C*). In other words, Sb atoms that sandwich the Mn atoms do not give rise to Dirac cones. In addition, there is a parabolic 2D band present along the $X \to R$ line at the zone boundary, which crosses the Fermi energy. The Fermi surface includes two parts: the Dirac cones and the electron pocket, all with 2D characteristic. The vertical lines (green) in Fig. 2*A* show the locations of four 2D Dirac cones. The low density of states at the Fermi level indicates that BaMnSb$_2$ is a semimetal, consistent with the high electrical resistivity shown in Fig. *1D*. An earlier calculation (15) using a similar procedure (*Materials and Methods*) but assuming a checkerboard-type AFM ground state reported an analogous Dirac-like and electron pocket at the Fermi energy.

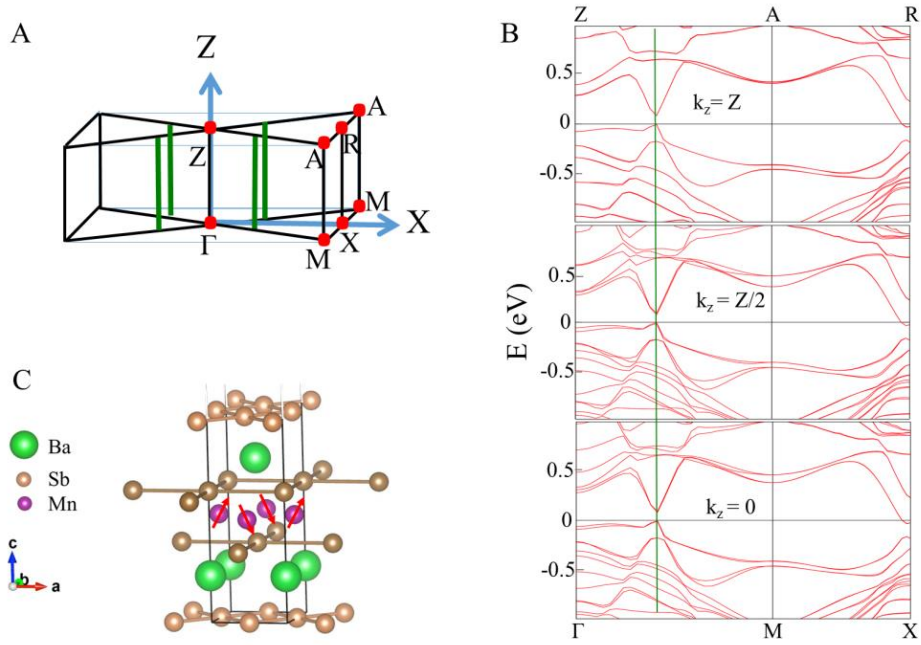

**Figure 2**: (A) Brillouin Zone of BaMnSb$_2$ (half). (B) Band structure within ± 1 eV from the Fermi energy for $k_z = 0$ (*Bottom*), Z/2 (*Middle*) and Z (*Top*). (C) Spin configuration of Mn.

Magnetically, the ground state corresponds the canted spin ordering of Mn with FM in the *ab* plane (with effective magnetic moment ~0.2$\mu_B$) and AFM coupling between planes (nearly zero magnetic moment) as illustrated in Fig. 2*C*. Such magnetic structure is consistent with the measured magnetic susceptibilities (Fig. 1*C*) and electrical anisotropy (Fig. 1*D*). Despite the obvious band degeneracy at the Dirac cones, a Weyl state could be established due to the magnetic structure, as suggested previously (2,12).

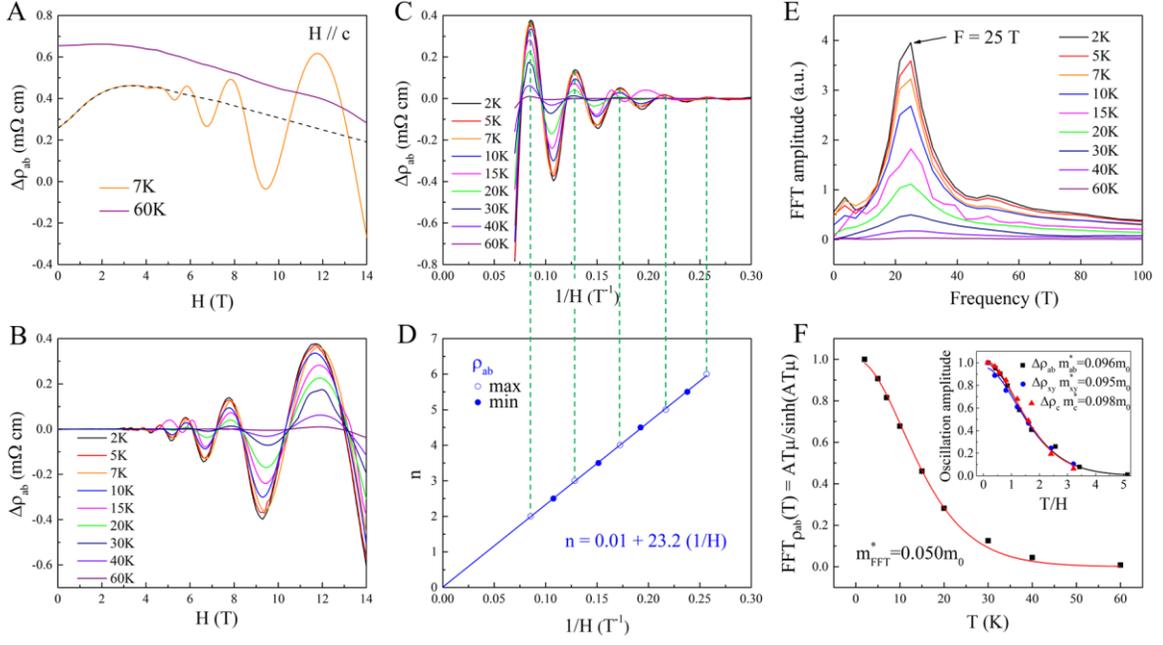

**Figure 3**: (A) Demonstration of the field dependence of $\rho_{ab}$ at 7 and 60 K. The dashed line represents the smooth background at 7 K. (B) Field dependence of oscillatory component $\Delta\rho_{ab}$ at indicated temperatures. (C) $\Delta\rho_{ab}$ plotted as a function of $H^{-1}$. (D) Landau level fan diagram obtained from $\Delta\rho_{ab}$. (E) Fast Fourier Transformation of oscillatory $\Delta\rho_{ab}$ at indicated temperatures. (F) Temperature dependence of FFT amplitude for $\Delta\rho_{ab}$. *Inset* shows normalized oscillation amplitude versus $T/H$ for $\Delta\rho_{ab}$, $\Delta\rho_{xy}$ and $\Delta\rho_{c}$ at the second LL.

To probe the character of the Weyl or Dirac state, we have measured $\rho_{ab}$, $\rho_c$, and Hall resistivity ($\rho_{xy}$) by applying magnetic field $H$ in the *ab* plane and *c* direction in a wide field and temperature range. All three quantities reveal Shubnikov-de Hass (SdH) oscillations for $H // c$ but not for $H // ab$. Fig. 3*A* shows the field dependence of $\rho_{ab}$ for $H // c$ at $T = 7$ and 60 K, demonstrating that $\rho_{ab}(H)$ depends nonmonotonically on $H$ with positive magnetoresistance (MR) [$\rho_{ab}(H) - \rho_{ab}(0)$] at low fields but negative MR at high fields. This is consistent with the weak ferromagnetism in the *ab* plane (Fig. 1*C*). Below

~ 50 K, MR$_{ab}$ clearly shows SdH oscillations at high fields. In Fig. 3*B*, we plot the field dependence of such oscillations $\Delta\rho_{ab}$ after subtracting a smooth background (dashed line in Fig. 3*A*). Although the amplitude increases with decreasing temperature and increasing field, the oscillatory pattern is the same. If we replot the data as $\Delta\rho_{ab}$ versus $H^{-1}$ as shown in Fig. 3*C*, the oscillations are at one frequency *F*. Different samples give the same frequency.

To understand the nature of SdH oscillations, Fast Fourier Transformation (FFT) analysis is carried out in the field range between 2.85 – 14 T. Fig. 3*E* shows the field dependence of FFT amplitude from $\Delta\rho_{ab}$, which reveals only a single frequency with *F* ~ 25 T. From the Onsager relation $F = (\Phi_0/2\pi^2)S_F$ (where $\Phi_0$ is the magnetic flux quantum), we obtain the extremal Fermi surface cross-sectional area normal to magnetic field $S_F$ ~ 0.24 nm$^{-2}$. It corresponds to the Fermi wave vector $k_F$ ~ 0.27(5) nm$^{-1}$ or $\frac{S_F}{(2\pi/a)^2} \approx 0.12(5)\%$ of the total area of the 2D Brillouin zone. *The finite $k_F$ indicates that the Fermi surface is not point-like, as seen in Fig. 2B*. The temperature dependence of the FFT can be used to determine the effective mass *m\** of electrons residing in this Fermi surface using the Lifshitz-Kosevich equation (16). We adopt the commonly used analysis procedures (10), where the FFT amplitude is fit to an equation $FFT_\rho(T) = \frac{AT\mu}{\sinh(AT\mu)}$, with $m^* = \mu m_0$ and $A = \frac{2\pi^2 k_B m_0}{e\hbar\bar{H}} = 3.085$ K$^{-1}$ (where $\mu$ is a constant, $m_0$ is the free electron mass, and $\bar{H}$ is the average magnetic field for our applied magnetic field range). The fit of this equation to our data, shown in Fig. 3*F*, yields $m^* = 0.050 m_0$, much smaller than *m\** for (Ca/Sr/Ba)MnBi$_2$ (6-10) and Sr$_{1-y}$Mn$_{1-z}$Sb$_2$ (13) and less than the value previously reported for BaMnSb$_2$ (11). However, it is larger than the value obtained for Cd$_3$As$_2$ (17, 18) using the fitting procedure described above. (We did the FFT, using data presented in ref. 18, and then calculated the effective mass from the FFT and the average magnetic field. The result was ~1/2 of the value obtained by analyzing the temperature dependence of a given LL, i.e., 0.02$m_0$.) If the same procedure is applied to the FFT amplitude obtained from the Hall resistivity oscillations (see below), the fit yields a similar effective mass $m^*$(Hall) = 0.057$m_0$. A more convincing procedure for determining the effective mass is to plot the amplitude of $\Delta\rho_{ab}$ at each LL as a function of $T$ from the data shown in Fig. 3*B*, where the magnetic field is fixed. The consistency of this procedure has been demonstrated for Cd$_3$As$_2$ (18), yielding an effective mass of 0.04$m_0$. Our data for the 11.7-T oscillations are plotted in Fig. 3*F*, *Inset*, with an effective mass of 0.096$m_0$, almost twice as large as the value obtained by averaging the FFT over the magnetic field range $\bar{H}$. As demonstrated in Fig. 3*F*, *Inset*, oscillation amplitude extracted directly from $\Delta\rho_{ab}$, Hall resistivity (Fig. 4*B*) and out-of-plane resistivity ($\Delta\rho_c$; Fig. 5*B*) yields the same effective mass (~0.097$m_0$). In other words, we obtain the same effective mass for every oscillation.

The nearly massless character of quasiparticles is consistent with what is expected from either Dirac or Weyl fermions. However, to be consistent it is important to check the Berry phase. Weyl or Dirac fermions are expected to possess so-called "zero mode" that does not shift with magnetic field, corresponding to a nontrivial Berry phase (19-21). To obtain information about the Berry phase $\phi_B$, one can apply the Lifshitz-Onsager quantization rule $S_n \frac{\hbar}{eH_n} = 2\pi \left(n + \frac{1}{2} - \frac{\phi_B}{2\pi}\right) = 2\pi(n + \gamma)$ to SdH oscillations. Here $S_n$ is the extremal cross-sectional area of the Fermi surface related to the *n*th Landau level; *e* is electron charge; $\gamma = \frac{1}{2} - \frac{\phi_B}{2\pi}$, which can be either 0 or 1. Because SdH oscillations are related to successive emptying of LLs as magnetic field is increased, we can experimentally determine *n* versus $H^{-1}$ from $\rho_{ab}$. Fig. 3D shows *n* as a function of $H^{-1}$ by assigning the maximum (peak) positions of $\Delta\rho_{ab}$ as a measure of the LLs. By constructing such a fan diagram, it is clear that oscillation at the highest magnetic field (11.7 T) is the n = 2 LL. The minimum (peak) positions of $\Delta\rho_{ab}$ are also collected and assigned to the half integers. The solid line in Fig. 3D is the fit of the data to the Lifshitz-Onsager relation, resulting in $\gamma \sim -0.01$. This value corresponds to $\phi_B \sim \pi$, indicating the *nontrivial* **Berry phase.**

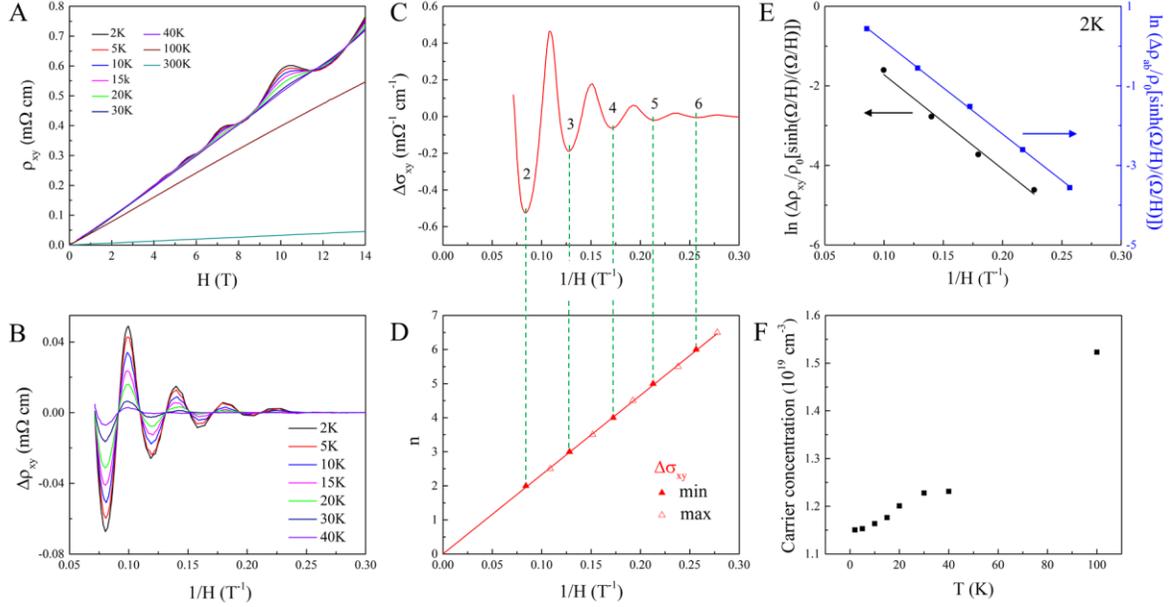

**Figure 4**: (A) Field dependence of Hall resistivity $\rho_{xy}$ at indicated temperatures. (B) Oscillatory Hall component $\Delta\rho_{xy}$ plotted as a function of $H^{-1}$. (C) Field dependence of oscillatory Hall ($\Delta\sigma_{xy}$) conductivities. (D) Landau level fan diagram constructed from $\Delta\sigma_{xy}(H)$ at 2 K. (E) Oscillatory $\Delta\rho_{ab}(H)$ and $\Delta\rho_{xy}(H)$ plotted as $ln\left(\frac{\Delta\rho}{\rho_0}\frac{sinh(\Omega/H)}{\Omega/H}\right)$ versus $H^{-1}$ ($\Omega = 2\pi^2 k_B m^* T/e\hbar$). (F) Temperature dependence of carrier concentration.

It has been pointed out (22) that the Landau level fan diagram (Fig. 3D) should be constructed from the electrical conductivity. The in-plane electrical conductivity is defined by $\sigma_{ab} = \rho_{ab}/(\rho_{ab}^2 + \rho_{xy}^2)$, meaning that the phase of oscillatory $\Delta\sigma_{ab}$ is not necessarily the same as $\Delta\rho_{ab}$. To confirm the nontrivial Berry phase, we have measured the Hall resistivity at various temperatures and fields. Fig. 4A displays $\rho_{xy}$ versus $H$. Several features are worth noting. First, $\rho_{xy}$ linearly depends on $H$ up to at least 14 T at each measured temperature, including the background when there are SdH oscillations below 50 K. This suggests that there is no anomalous Hall effect due to magnetism. Second, the slope $R_H$ is positive, reflecting the hole-dominated electrical conduction. Although it remains constant below 50 K, $R_H$ tends to decrease with increasing $T$ at higher temperatures (Fig. 4A). This is different from previous work for the same material (11) but similar to that observed in the Dirac semimetal $Cd_3As_2$ (17). Third, similar to $\rho_{ab}$, there are SdH oscillations in $\rho_{xy}$ below 50 K. The field and temperature dependence of the oscillatory component $\Delta\rho_{xy}$ is displayed in Fig. 4B, plotted as $\Delta\rho_{xy}$ versus $H^{-1}$. As shown in Fig. 3F, *Inset*, the temperature and field dependence of the oscillation amplitude is the same as $\Delta\rho_{ab}$.

The phase of $\Delta\rho_{xy}$ oscillations shifts approximately π/2 from $\Delta\rho_{ab}$. This is the same as observed in $Cd_3As_2$ (17) and $Bi_2Te_3$ (23). With $\rho_{ab}$ and $\rho_{xy}$, we calculate $\Delta\sigma_{xy}$ via $\sigma_{xy} = \rho_{xy}/(\rho_{ab}^2 + \rho_{xy}^2)$. Fig. 4C plots $\Delta\sigma_{xy}(H)$ at 2 K as a function of $H^{-1}$. We index $n$ with the minimum (valley) positions of these quantities as recommended in ref. 22 and plot $n$ as a function of $H^{-1}$ in Fig. 4D. For precise fitting, the maximum (peak) positions of these quantities are also collected and assigned to the half integers. The solid line in Fig. 4D is the fit of the data to the Lifshitz-Onsager relation, resulting in $\gamma \sim 0.006$ for $\Delta\sigma_{xy}$. This corresponds to $\phi_B \sim \pi$, again the nontrivial Berry phase.

The quantum mobility of Weyl or Dirac fermions can be estimated from the damping of SdH oscillation amplitude. The oscillation amplitude $\Delta\rho$ is related to the quantum mobility $\mu_q$ by $\Delta\rho = \rho_0 e^{-\pi/\mu_q H} \frac{2\pi^2 k_B m^* T/e\hbar H}{sinh(2\pi^2 k_B m^* T/e\hbar H)}$. As shown in Fig. 4E, one can obtain $\mu_q$ from the linear fit of $ln\left(\frac{\Delta\rho}{\rho_0} \frac{sinh(2\pi^2 k_B m^* T/e\hbar H)}{2\pi^2 k_B m^* T/e\hbar H_n}\right)$ vs. $H^{-1}$ for both $\Delta\rho_{ab}$ and $\Delta\rho_{xy}$ at 2 K, which results $\mu_q \sim$ 1325 – 1349 cm$^2$ V$^{-1}$ s$^{-1}$ and the quantum relaxation time $\tau_q \sim$ 7.2 – 7.4×10$^{-14}$ s ($\mu_q = e\tau_q/m^*$). This $\mu_q$ is higher than that of SrMnBi$_2$ (250 cm$^2$ V$^{-1}$ s$^{-1}$) (8), Sr$_{1-y}$Mn$_{1-z}$Sb$_2$ (570 cm$^2$ V$^{-1}$ s$^{-1}$) (13), and BaMnSb$_2$ (1280 cm$^2$ V$^{-1}$ s$^{-1}$) (11) but much smaller than that obtained in Cd$_3$As$_2$ (24,25). The transport mobility $\mu_t$ obtained from Hall resistivity data shown in Fig. 4A at 2 K is estimated to be 2200 cm$^2$ V$^{-1}$ s$^{-1}$, higher than the quantum mobility. Using the Drude model, we estimate the carrier concentration from Hall coefficient (= $\rho_{xy}/H$) for each temperature. Fig. 4F shows the temperature dependence of carrier concentration below ~ 100 K, which is on the order of 10$^{19}$ cm$^{-3}$. This value is about two orders higher than that of Cd$_3$As$_2$ (26), consistent with the metallic $\rho_{ab}$ character.

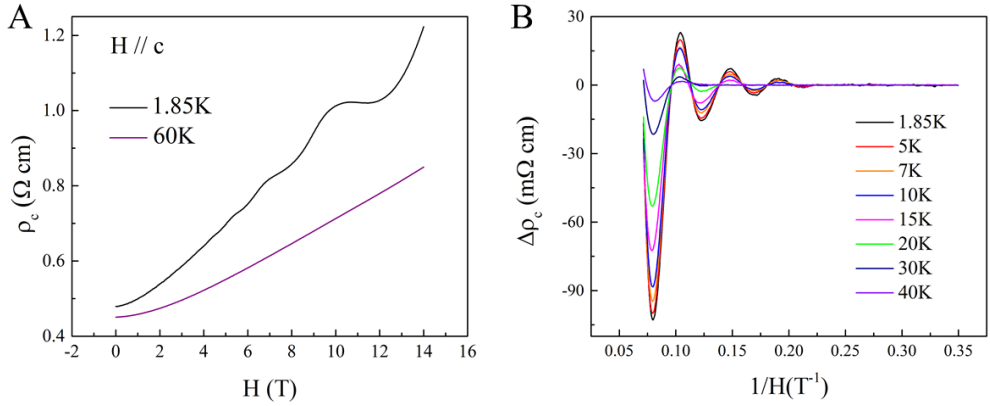

**Figure 5**: (A) Demonstration of the field dependence of $\rho_c$ at 1.85 and 60 K. (B) Field dependence of oscillatory component $\Delta\rho_c$ at indicated temperatures.

Given that $\rho_c \gg \rho_{ab}$ (Fig. 1D), the question is, can quantum oscillation be observed in $\rho_c$? Fig. 5A shows field ($H \parallel c$) dependence of $\rho_c$ at 1.85 K (below the bump) and 60 K (above the bump). The positive MR is consistent with the antiferromagnetic nature along the $c$ direction. Similar to $\rho_{ab}$, SdH oscillations are clearly seen at low temperatures ($T < 50$ K). By subtracting the background, we obtain $\Delta\rho_c$ for various temperatures. As shown in Fig. 5B, the amplitude of oscillatory $\Delta\rho_c$ increases with decreasing temperature, the same as $\Delta\rho_{ab}$ and $\Delta\rho_{xy}$ (Fig. 3F, *Inset*). The obvious difference is the oscillations of $\Delta\rho_{ab}$ and $\Delta\rho_c$ have a $\pi$ phase shift: when $\Delta\rho_{ab}$ reaches a maximum, $\Delta\rho_c$ is at a minimum and vice versa. Previous work showed the same behavior in BaMnSb$_2$ (11) but not in Sr$_{1-y}$Mn$_{1-z}$Sb$_2$ (13). This should be a good material system for studying the Berry phase effect on electronic properties (27).

In summary, we have successfully synthesized high-quality BaMn$_{1-\delta}$Sb$_2$ single crystals with small Mn deficiency ($\delta < 0.05$). Both experimental and theoretical investigations indicate that BaMn$_{1-\delta}$Sb$_2$ is a magnetic semimetal with ferromagnetism in the *ab* plane but AFM coupling along the *c* direction. Experimentally, we observe SdH oscillations in the in-plane, out-of-plane, and Hall resistivities when a magnetic field is applied along the *c* direction below ~50 K. Our data unambiguously demonstrate that such oscillations result from nearly massless quasiparticles and include an additional phase, i.e., *a nontrivial π Berry phase*. This can only be realized if electron and hole pockets are crossing to form Dirac or Weyl points. Furthermore, the small but finite wave vector indicates that such points are not exactly at the Fermi level, implying type-II Dirac/Weyl character. Although the broken time-reversal symmetry due to magnetic ordering is a necessary condition, it is not sufficient to have Weyl splitting. For example, a magnetic compound, particularly AFM ordered, breaks time-reversal symmetry. However, other magnetic symmetry may exist, which can prevent Weyl splitting. Antiferromagnetic EuMnBi$_2$ is an example, which shows no evidence for Dirac/Weyl fermions at low magnetic filed (28,29). The application of high magnetic field results in spin flop, thus forming Dirac fermions due to spin degeneracy lifting in EuMnBi$_2$ (29). Our BaMnSb$_2$ with FM ordering in the *ab* plane may be the key to realize Weyl fermions, and work assures its important placement in the small Weyl material family. Unfortunately, standard density functional theory (DFT)/generalized gradient approximation (GGA) + U calculations do not capture the effect of broken time-reversal symmetry due to magnetic ordering to predict a Weyl semimetal (15). The authors of ref. 15 had to break the lattice symmetry to obtain a Weyl semimetal for BaMnSb$_2$. A possibility for this is a lack of non-local Hartree-Fock exchange, which DFT/GGA + U does not take into account, but a hybrid-DFT approach could address this discrepancy.

**Materials and Methods**

Barium pieces (99+% Alfa Aesar), manganese powder (99.5% Alfa Aesar), and antimony powder (99.5% Alfa Aesar) were used for preparation of $BaMnSb_2$. Elements were weighed out using a molar ratio of 1:1:2 (Ba:Mn:Sb), placed into a alumina crucible, and loaded into a fused silica tube which was evacuated (~10 millitorr) and sealed. The reaction tube was heated to 650 °C at 150 °C/h, held at 650 °C for 1 h, further heated to 750 °C at 50° C/h, held at 780 °C for 1 h, and cooled down to room temperature by turning off the power. The sample was grounded and sealed again in quartz tube under vacuum. Single crystals were grown with the growth speed of 3 mm/hr in the floating-zone furnace. The structural (Fig. S1) and composition characterization (Table S1) is described in the *Supporting Information*.

The relativistic electronic structure calculations were performed using the periodic density functional theory program Vienna *ab initio* simulation package (VASP), version 5.2.12 (30,31). The Kohn-Sham equations were solved using the projector augmented wave approach (32) and a plane wave basis with a 387.187 eV energy cutoff. The GGA exchange correlation functional of Perdew, Burke, and Ernzerhof was used (33). Brillouin zone sampling was done using the Monkhorst-Pack method (34) with a *k*-point mesh of 8×8×4. The electronic self-consistent field was deemed converged if the total energy between successive iterations differed by less than 0.0005 eV per atom. Full geometry relaxation was performed using a conjugate gradient algorithm and were deemed converged if a subsequent optimization step agreed with the before within 0.001 eV per atom.

**Acknowledgements**

We are grateful for valuable discussion with Zhong Fang and Wei Ku. The experimental work is supported by NSF through Grant Number DMR-1504226. The computational work conducted by W.A.S. is supported by an allocation of computing time from the Louisiana State University High Performance Computing Center.

**Supporting Information to the Manuscript "Nontrivial Berry Phase in Magnetic BaMnSb$_2$ Semimetal"**

Experimental investigations were performed in single crystal samples. After the growth in a floating-zone furnace, single crystals were crushed to powder for X-ray diffraction (XRD) measurement. Fig. S1 shows the XRD pattern, where all peaks can be indexed by a single phase with *I4/mmm* symmetry with lattice parameters $a = b = 4.556$ Å and c = 24.299 Å.

The chemical composition of single crystals is measured via the Wavelength Dispersive Spectroscopy (WDS) technique. Table S1 lists the elements and corresponding atomic ratio from 9 spots. While there is slight composition variation, the result indicates that the atomic ratio of Ba : Mn : Sb ~ 1 : 1 : 2 with small amount of Mn deficiency (~5%).

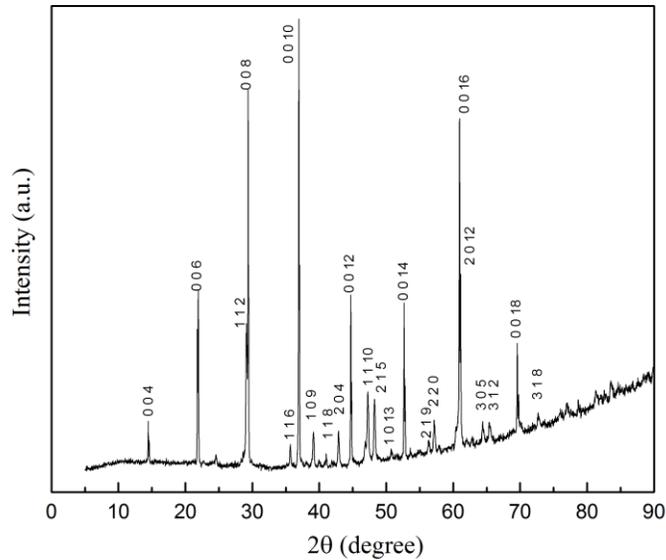

Figure S1: X-ray diffraction pattern of BaMnSb$_2$ powder by crushing single crystals.

Table S1. Composition analysis via WDS measurement

| Element | Point | | | | | | | | | |
| --- | --- | --- | --- | --- | --- | --- | --- | --- | --- | --- |
| | 1 | 2 | 3 | 4 | 5 | 6 | 7 | 8 | 9 | Average |
| Ba (atom%) | 26.8719 | 24.3470 | 25.4885 | 24.4076 | 24.9969 | 25.0404 | 26.1468 | 24.9850 | 24.8816 | 25.2406 |
| Mn (atom%) | 23.7523 | 24.2355 | 23.2794 | 24.3455 | 23.6619 | 24.6668 | 24.5749 | 23.4544 | 23.8044 | 23.9750 |
| Sb (atom%) | 49.3758 | 51.4175 | 51.2321 | 51.2468 | 51.3412 | 50.2928 | 49.2783 | 51.5607 | 51.3140 | 50.7843 |